\documentclass[aps,twocolumn,superscriptaddress,showpacs]{revtex4}

\usepackage{epsf,latexsym}
\usepackage{times}

\newcommand{\ket}[1]{| #1 \rangle}
\newcommand{\braket}[2]{\left \langle #1 | #2 \right\rangle}

\newcommand{\be}{\begin{equation}}
\newcommand{\ee}{\end{equation}}

\def\CC{{\rm\kern.24em \vrule width.04em height1.46ex depth-.07ex
    \kern-.30em C}}
\def\P{{\rm I\kern-.25em P}}

\def\bbbone{{\mathchoice {\rm 1\mskip-4mu l} {\rm 1\mskip-4mu l}
{\rm 1\mskip-4.5mu l} {\rm 1\mskip-5mu l}}}

\def\bbbc{{\mathchoice {\setbox0=\hbox{$\displaystyle\rm C$}\hbox{\hbox
to0pt{\kern0.4\wd0\vrule height0.9\ht0\hss}\box0}}
{\setbox0=\hbox{$\textstyle\rm C$}\hbox{\hbox
to0pt{\kern0.4\wd0\vrule height0.9\ht0\hss}\box0}}
{\setbox0=\hbox{$\scriptstyle\rm C$}\hbox{\hbox
to0pt{\kern0.4\wd0\vrule height0.9\ht0\hss}\box0}}
{\setbox0=\hbox{$\scriptscriptstyle\rm C$}\hbox{\hbox
to0pt{\kern0.4\wd0\vrule height0.9\ht0\hss}\box0}}}}

\def\bbbz{{\mathchoice {\hbox{$\sf\textstyle Z\kern-0.4em Z$}}
{\hbox{$\sf\textstyle Z\kern-0.4em Z$}}
{\hbox{$\sf\scriptstyle Z\kern-0.3em Z$}}
{\hbox{$\sf\scriptscriptstyle Z\kern-0.2em Z$}}}}

\newcommand{\putfig}[2]{$$\leavevmode\hbox{\epsfxsize=#2 cm
   \epsffile{#1.eps}}$$}

\begin{document}
\title{All-Electrical Quantum Computation with Mobile Spin Qubits}

\author{A.E.~Popescu}
\affiliation{Department of Engineering, University of Cambridge, Cambridge CB2 1PZ, UK}
\author{R.~Ionicioiu}
\affiliation{Institute for Scientific Interchange (ISI), Villa Gualino, Viale Settimio Severo 65, I-10133 Torino, Italy}

\begin{abstract}
We describe and discuss a solid state proposal for quantum computation with mobile spin qubits in one-dimensional systems, based on recent advances in spintronics. Static electric fields are used to implement a universal set of quantum gates, via the spin-orbit and exchange couplings. Initialization and measurement can be performed either by spin injection from/to ferromagnets, or by using spin filters and mesoscopic spin polarizing beam-splitters. The vulnerability of this proposal to various sources of error is estimated by numerical simulations. We also assess the suitability of various materials currently used in nanotechnology for an actual implementation of our model.
\end{abstract}


\maketitle

\section{Introduction}

During the last decade, quantum computing has emerged as a theoretically viable option that surpasses classical methods from the point of view of computational power. The algorithm discovered by Shor \cite{Shor} came to show that potential advantages, such as the exponential speed-up, can be used to solve problems of practical interest \cite{QC1,QC2,QC3}. Since then, numerous approaches for implementing quantum information processing have been proposed, ranging from optical to chemical environments and methods. Although benefiting from considerable theoretical and technological knowledge, solid-state systems have been rather late contenders in the race for potential quantum computing implementations. At present, the most advanced solid-state proposals are those using superconducting qubits (see for example \cite{sc5,sc6} and references therein).

Although difficult to manipulate and sensitive to various sources of decoherence, the degrees of freedom of a single particle can act as a natural depositary for quantum information. Some research has been directed towards the investigation of quantum computing with charge qubits, where information is encoded in the orbital degrees of freedom of an electron \cite{charge1,charge3}. The main problem of all these proposals is a very short charge decoherence time (of the order of picoseconds). This, plus less obvious considerations, such as a non-Markovian nature of the decoherence induced by Fermi sea effects, which may impede quantum error correction in such systems \cite{DiVi1}, leads to serious questioning of the suitability of a single electron charge as a proper qubit.

Another promising group of models, although not yet demonstrated experimentally, rely on using the spin of a single particle (electron or nucleus) as a qubit. The first proposal which takes advantage of both the impressive stability of nuclear spins (decoherence times of up to hours for P donors in Si) and the electron capacity to mediate interactions was reported by Kane \cite{Kane1}. Another implementation based on similar principles has been proposed in Ref.~\cite{Smet}. The manufacturability and scalability of this group of proposals still depend on further technological advances, although first steps have already been taken \cite{Kane2}.

In this paper we focus on encoding quantum information in the spin of a single electron. The seminal paper of Loss and DiVincenzo \cite{LossDiVi} defines the qubit as the spin of an electron located in a quantum dot, thus counting on dephasing times around 100\,ns (see for instance \cite{Engel} and references therein). A complete set of one qubit gates is implemented using local magnetic fields with different orientations, which lead to Zeeman splitting in the quantum dot. A drawback of this proposal is the requirement for local control of the magnetic fields, but this can be overcome by moving the electrons to and from areas where desired magnetic fields are easily obtained (although at the expense of an increased number of operations, i.e., repeated applications of the ${\sf SWAP}$ gate). The two-qubit interaction responsible for the crucial creation of entanglement is the exchange coupling between electrons in adjacent quantum dots, mediated by electric potentials. It was shown that by turning on the exchange for an appropriate time, the universal gate $\sqrt{\sf SWAP}$ can be implemented. Related proposals include more realistic physical setups, either by relaxing the technological constraints \cite{Privman}, or by describing the physical system involved with increased accuracy \cite{multiple}. States outside the computational space are avoided only if the electric and magnetic fields applied to each dot are turned on and off adiabatically \cite{Golovach}. A variation to this proposal, detailed in \cite{DiVi3}, has introduced the concept of {\em encoded universality}, thus eliminating the constraint of locally controllable magnetic fields. In this approach, an universal set of gates is implemented solely by exchange coupling, with the added advantage of an improved protection against decoherence \cite{ex_DFS}. Various other electron spin implementations have been proposed in \cite{Burk,Imam,Vrij,Bandy,Pazy,Levy,Benjamin,STM,Khitun,Salis,Bose,Zhou,Friesen1,Friesen2,Wu,Feng}.

The computational function of a quantum computer has to be complemented by the transport of quantum information among various parts of the circuit \cite{DiVi2}. In models that use static qubits, this task is implemented using a chain of quantum dots; the state of the qubit, rather than the qubit itself, is transported along such chains via exchange interactions between neighboring quantum dots, introducing a number of extra operations without any computational function. A different approach was proposed by Barnes {\em et al.} in \cite{Barnes}. Mobile qubits are used, rather than static ones; instead of applying a succession of electric and magnetic fields to a quantum dot, the qubit itself is moved around the quantum circuit, passing through the gates, implemented by predefined areas with static electric and magnetic fields. This eliminates the requirement of ultrafast switching of electric and magnetic fields (on picoseconds time scales), while imposing extra constraints regarding coherence and synchronization. In the Barnes {\em et al.} proposal the electrons are extracted from a two-dimensional electron gas (2DEG) and moved through the gates using a surface acoustic wave (SAW). One-qubit gates are implemented with static magnetic fields and entanglement is created by exchange coupling.

In this paper we discuss the feasibility of a new type of implementation for quantum logic circuits, using mobile spin qubits manipulated solely by static electric fields. Quantum computing with mobile electron spins is a natural extension of spintronics, since it can use all theoretical and experimental developments in this field. Methods for preparation, manipulation and measurement are detailed in Section II, whereas Section III is dedicated to the simulation of a typical two-qubit gate. Performance criteria, such as scalability, are assessed in Section IV. In Section V, we examine various materials (semiconductor heterostructures, Si, carbon nanotubes) which could be used for practical implementations of the scheme proposed here. The main advantages and drawbacks of our approach are summarized in Section VI.

\section{The model}\label{model}

\subsection{Spin-coherent transport}

One of the main results that prompted this study was the experimental demonstration of spin-coherent transport over distances exceeding 100$\mu$m in semiconductor substrates. In Ref.~\cite{optic1}, a spin-coherent current was induced by exciting the substrate with polarized light, then transported along a sample of GaAs over distances larger than 100$\mu$m (at 1.6\,K). It is conceptually possible to narrow the sample, so that only one electron will pass at a time, without significantly reducing the coherence length. Ballistic transport, where the mean free path of the electron spin is much larger than the dimension of the device, has also been demonstrated in one-dimensional, 5$\mu$m long samples of GaAs/AlGaAs \cite{ball_spin}. More recently, ballistic pure spin currents have been observed in ZnSe \cite{spin_currents_1} and GaAs/AlGaAs quantum wells \cite{spin_currents_2}.

The field of spin-coherent transport in one-dimensional nanowires is open to rapid progress and improved coherence lengths can be expected in the future, especially when the fabrication of controlled carbon nanotubes (CNTs) will become available on a larger scale (for more details see Section V.C).  

\subsection{Definition of the qubit}

We define the qubit as the spin state of an electron moving in a quantum wire, where: ${\ket{0}}~\equiv~$spin-up and ${\ket{1}}~\equiv~$spin-down (``flying qubit''). The archetypal flying qubit is a photon and its two polarization states, and thus our qubit can be seen as the mesoscopic analog of an optical qubit.

The electrons propagate along one-dimensional (1D) quantum wires situated in the $xy$-plane and are confined in the $z$-direction. Between the gates the qubit is described by the free Hamiltonian:
\begin{equation}
H_0= \frac{\bf p^2}{2m^*}+ V_0({\bf r})
\label{H0}
\end{equation}
For a 2DEG, we can take $V_0({\bf r})$ as parabolic in the (in plane) direction perpendicular to the wire and a rectangular potential well in the growth $z$-direction. From now on we will neglect any contribution in the $z$-direction and treat it as a two-dimensional problem in the $xy$-plane.

We assume that the electrons are injected at time intervals large enough so that intra-wire interactions (i.e., the Coulomb coupling between successive electrons in the same wire) are negligible.

\subsection{Initialization}\label{initialisation}

The approach to spin preparation used in Ref.~\cite{LossDiVi}, namely slow cooling of electron spins in their respective sites before the computation starts, is also suitable for our proposal. Recent advances in spintronics offer a number of alternatives for the initialization of a mobile spin qubit, using methods ranging from optical \cite{optic1,optic2} to electrical. However, the options we prefer are based on purely electric methods, such as spin filters and spin injection from ferromagnets to semiconductors.

Spin filters in various designs have been proposed, based either on electron transport through layers of magnetic materials \cite{filter1,filter2,filter3,filter4,filter5}, or on the Rashba effect \cite{filter6,filter7,filter8,filter9}. A simple and very effective spin filter can be implemented by using a quantum dot with a special band structure \cite{filter_dot}, although this result has not been demonstrated experimentally yet (and could require the presence of a controllable, albeit global, magnetic field). Spin filters and spin pumps have been recently demonstrated experimentally \cite{spin_filter, spin_pump}. A spin pump which uses only electric currents has been examined in Ref.~\cite{spin_pumpE}.

A mesoscopic spin polarizing beam splitter (PBS) \cite{pbs} can be used as both a spin preparation and a spin measuring device. An incident unpolarized spin current in input 0 is split into two polarized outputs: a spin up (down) will always exit (i.e., with unit probability) in the 0 (1) output. As a spin measuring device it has the advantage of conserving the number of particles between input and output (a spin filter will absorb some of the spins), so no spins are lost.

Injection of spin-coherent currents from ferromagnetic substrates into semiconductors benefits from very promising experimental results in spintronics (for a review of the subject see Ref.~\cite{Wolf}). Although the efficiency of injection was initially low, due to the large contact resistance between two different materials, there are encouraging results for high-efficiency injection from a magnetic to a non-magnetic semiconductor with similar band structure \cite{injection1} or by using a magnetic semiconductor as a spin aligner \cite{injection2}. Recent results for Fe/GaAs/Fe, Fe/ZnSe/Fe (001) junctions show almost ideal injection efficiencies \cite{injection3,injection4}. The possibility of spin injection at room temperature was explored in \cite{injection5}, although the spin polarizations achieved are still modest. It was also predicted in \cite{Flatte2} that high electric fields can enhance spin injection from a ferromagnetic metal into a semiconductor, helping to overcome the drawback of a high contact resistance.

\subsection{Read-out}\label{readout}

One of the most delicate problems for quantum computation with spin qubits is the read-out step, since measuring a single spin is a notoriously difficult task. A number of methods have been proposed, based on optical principles \cite{optic1,injection2} (where the electrons are allowed to recombine and the polarization of the emitted light is measured), scanning tunneling microscopy \cite{STM}, or magnetic resonance force microscopy \cite{MRF}. Another method for single spin detection is based on converting the spin into charge  (for instance by a procedure similar to the one described in Ref.~\cite{Pazy}), which can be subsequently measured using a single electron transistor (SET).

Using mobile electrons rather than static ones allows for an easier measurement method well suited for our set-up. A spin filter allows only electrons with a specific spin polarization to be transported across it, and will absorb the others. The presence of an electron -- meaning that the measured state was not filtered out -- can be detected using a SET. Thus, the problem of measuring a single spin is transfered, if the experiment is repeated a number of times, into counting statistics for a charge current coming out of the filtering device. The simplest design for a spin filter is a ferromagnetic lead, with the band structure matched to the one of the material used to build the actual quantum circuit. Solutions similar to the case of spin injection described above can be designed in order to improve the efficiency of this detector. A mesoscopic spin polarizing beam splitter (PBS) \cite{pbs} can also be used as a measuring device with a potentially large efficiency.

Another advantage of our proposal is that the measuring device is physically separated from the qubit. Mobile qubits arrive at the detector at the end of the calculation. For static qubits, the measuring apparatus is usually situated close to the qubit and is coupled to the qubit only at the end of the computation. This proximity can introduce extra decoherence, e.g., in the case of a superconducting qubit having the readout SQUID close to the qubit. An exception is the optical readout of a static qubit.

\subsection{Quantum gates}

A universal set of quantum gates include, for example, all one-qubit rotations and a two-qubit entangling gate, such as ${\sf CNOT}$ or $\sqrt{\sf SWAP}$ \cite{univ1,univ2,LossDiVi}. In order to implement these transformations it is important to take advantage of the natural interactions that couple to the electron spin in a solid-state environment. In the present approach we choose to control the qubit using only static electric fields in order to avoid the more delicate manipulation of magnetic fields. It should be stressed that stray magnetic fields will always be present in the quantum circuit; their effects should be included in detailed calculations or numerical simulations, either by considering supplementary terms in the Hamiltonian of the system or by treating them as extra sources of decoherence. A range of possible interactions leading to an universal set of gates (by themselves or in combinations) are discussed below. The suitability of each interaction for particular practical implementations will be addressed in Section V.

\subsubsection{Single qubit gates: the spin-orbit coupling}

Even in the absence of external magnetic fields, spin rotation of a mobile electron can still be achieved with the so-called spin-orbit interaction. An electron moving with velocity ${\bf v}$ in a region with a {\em static} electric field ${\bf E}$ will see an effective magnetic field ${\bf B \sim v \times E}$ which couples to its spin. The spin-orbit Hamiltonian due to this coupling is $H_{so}\sim \vec{\sigma} \cdot ({\bf k \times E})$, where {\bf k} is the electron wave-vector and $\vec{\sigma}= (\sigma_x,\sigma_y,\sigma_z )$ is the vector of Pauli matrices. The Hamiltonian $H_{so}$ contains the necessary ingredients for implementing spin rotations around two independent axes and therefore an arbitrary single-qubit operation.

We now discuss two possible architectures. For a 2DEG situated in the $xy$-plane the spin-orbit coupling is given by the Rashba effect \cite{Rashba}, with the following Hamiltonian \cite{Rashba3}:
\begin{equation}
H_{Rashba}= \alpha (\sigma_x k_y - \sigma_y k_x)
\label{H_rashba}
\end{equation}
where $\alpha$ is the spin-orbit coupling. The Rashba interaction arises due to the strong interface electric field in the growth $z$-direction. The coupling $\alpha$ depends on the structure of the specific material used and can be controlled by a surface electric field applied by top/bottom gates \cite{Rashba1, Rashba2}. An approximate dependence of $\alpha$ on the electric field $E$ is the following \cite{Bandy1}:
\begin{equation}
\alpha\simeq\frac{e\hbar}{2(m^* c)^2} E
\label{alpha}
\end{equation}
($m^*$ is the effective electron mass). However, equation (\ref{alpha}) does not include material dependent effects, such as build-in fields or variations in the electron density and mobility, hence an experimental value is preferred whenever available.

In this case spin rotations ${\sf R_x}(\theta)\equiv e^{i\theta \sigma_x}$ and ${\sf R_y}(\theta)\equiv e^{i\theta \sigma_y}$ can be implemented by controlling the propagation direction of the particle. Thus, if the electron is propagating along the $x$ ($y$)-axis with wave-vector $k_x$ ($k_y$), the Rashba-active region will induce a spin rotation $\sf R_y$ ($\sf R_x$), as in Fig.\ref{su2}(b).

A second (equivalent) configuration is possible if both top/bottom and lateral gates are experimentally feasible and if the spin-orbit coupling corresponding to two directions are comparable in strength. In this case the electron can move along a single direction, say $x$, and the spin-orbit Hamiltonian reduces to:
\begin{equation}
H_{so;x}= k_x (\alpha_y \sigma_z- \alpha_z \sigma_y)
\label{H_Rashba_x}
\end{equation}
where $\alpha_{y,z}$ is the spin-orbit coupling which includes the effect of the applied field $E_{y,z}$. Now spin rotations $\sf R_y$ and $\sf R_z$ are enacted by controlling the electric fields $E_z$ and $E_y$, respectively (see Fig.\ref{su2}(a)).

A straightforward calculation shows that the evolution under the Hamiltonian $k_x \alpha_y \sigma_z$ gives a spin rotation angle $\theta_z= \alpha_y m^*L_x/\hbar^2$, where $L_x$ is the gate length in the $x$-direction; similar results can be obtained for the other two rotations. The electric field can be controlled by applying a potential on a local capacitor. All single-qubit transformations can be implemented in maximum three steps using spin rotations around two axes \cite{univ1}.

A very important property of the Rashba one-qubit gate is its non-dispersivity. It can be seen from the above transformation that the rotation angle $\theta_z= \alpha_y m^*L_x/\hbar^2$ depends only on the local gate parameters (its length $L_x$ and the strength of the applied field $E$, through $\alpha$) and not on the energy of the incoming electron (this is correct if the interband coupling is negligible, which is true if the channel width $w \ll \hbar^2/\alpha m^*$ \cite{datta_das}).

We can estimate the length $L$ of the Rashba region necessary for a rotation angle $\theta= \pi$ as $L= 116\,$nm in InAs ($\alpha= 4\times 10^{-11}$eVm \cite{Rashba2}) or $L= 500\,$nm in InGaAs/InAlAs ($\alpha= 0.93\times 10^{-11}$eVm \cite{Rashba1}). These values give a figure of 200 to 900 single-qubit gates for a spin coherence length of 100 $\mu$m. It is important to note that a precise control of the rotation angle with the surface electric field was achieved, therefore it is possible to implement the continuous set of rotations necessary for universal quantum computation.

We define the error of a single gate as:
\begin{equation}
\epsilon=1- \min\limits_{\ket{\psi_{in}}} \left| \braket{\psi_{out}}{\psi_{out}^0} \right|
\end{equation}
where $\ket{\psi_{out}^0}$ is the desired output state, $\ket{\psi_{out}}$ is the real output (which includes all the gate errors) and the minimum is taken over all possible input states $\ket{\psi_{in}}$.

Let $\ket{\psi_{in}}= \cos\delta\ket{0}+ e^{i\gamma}\sin\delta \ket{1}$ be a general one-qubit state. For a rotation around the $y$-axis with angle $\theta$, the error can be written as:
\begin{eqnarray}
\nonumber
\epsilon_y= 1- \frac{1}{2} \min\limits_{\delta,\gamma} \left| \sin 2\delta\, \{ \cos(\theta-\theta_0)+ \right. \\
\left. +[e^{-i\gamma}-e^{i\gamma}\sin(\theta-\theta_0)~] \} \right|
\end{eqnarray}
where $\theta_0$ is the desired value of the rotation angle, while a rotation around the $z$-axis with angle $\theta$ introduces the following error:
\begin{equation}
\epsilon_z= 1-\min\limits_{\delta}\left| \cos^2\delta\, e^{i(\theta-\theta_0)}+ \sin^2\delta\, e^{-i(\theta-\theta_0)} \right|
\end{equation}

The maximum error occurs for input states of the form $\cos\delta\,\ket{0}+\sin\delta\,\ket{1}$ for $\sf R_y$ and $(\ket{0} + e^{i\gamma} \ket{1})/ \sqrt{2}$ for $\sf R_z$ and is the same in both cases:
\begin{equation}
\epsilon_y^{max}= \epsilon_z^{max}= 1- | \cos(\theta-\theta_0)\, |
\end{equation}

For instance, a variation of $10\%$ in the gate length will induce an error in the rotation angle of approximately $5\%$. Since the rotation angle $\theta_z\sim \alpha_y L_x$ depends only on the gate length and the electric field (through $\alpha$), fabrication errors in the gate length can be compensated by local adjustment of the electric field. 

We mention that the Rashba effect was also used for providing spin splitting in a static qubits proposal (electrons in self-assembled quantum dots) in Ref.~\cite{Bandy}.

\subsubsection{Two qubit gates: the exchange coupling}

The Hamiltonian for two spins interacting via isotropic exchange can be written as \cite{LossDiVi}:
\begin{equation}
H_{exchange}= J(t)~{\bf S_1}\cdot{\bf S_2}
\label{Hex}
\end{equation}
where $J(t)$ depends on the overlap of electron (orbital) wave-functions, and is zero if one of the electrons is outside the two-qubit gate. The exchange coupling $J$ can be turned on either by reducing the inter-wire distance or by lowering the potential between the two electron sites \cite{Burk}. Electric control has the advantage of allowing the gate to be switched on and off by external potentials, thus enabling the programming of our quantum circuit.

Since ${\bf S_1}\cdot{\bf S_2}= (2 U_{Swap}- \bbbone)/4$, the time evolution of the system under the exchange Hamiltonian (\ref{Hex}) is given by:
\begin{equation}
U_{exchange}=e^{i\beta/4}\exp(-i\beta U_{Swap}/2)
\end{equation}
where $U_{Swap}= {\rm diag} (1, \sigma_x, 1)$ is the matrix of a ${\sf SWAP}$ gate, and
\begin{equation}
\beta= \frac{1}{\hbar}\int_{t_1}^{t_2}J(t)dt
\end{equation}
(both electrons are assumed to be in the gate region between times $t_1$ and $t_2$). For $\beta= \pi$, the spin states of the two interacting electrons are swapped; $\beta= \pi/2$ corresponds to the universal gate $\sqrt{\sf SWAP}$ (up to an overall phase). The usual ${\sf CNOT}$ gate can be obtained using two $\sqrt{\sf SWAP}$ gates and three one-qubit gates \cite{LossDiVi}.

In the presence of anisotropy, an extra term appears in the exchange Hamiltonian; its effect on the gate accuracy has been addressed in \cite{aniso_exchange}. As already mentioned, the exchange interaction becomes universal by using a particular encoding \cite{DiVi3}, and moreover it can be protected against collective decoherence using decoherence-free subspaces (DFS) \cite{ex_DFS}.

\section{Simulation of a two-qubit gate}

In this section we assess the effect of various parameters on the exchange coupling $J$ through numerical simulations. Gate accuracy can be affected by two types of parameters: (i) geometrical, like gate length, inter-wire distance, barrier height; and (ii) dynamical, e.g., energy difference between the two electrons and the time lag between them (this is due to synchronization errors when one electron arrives at the gate before the other).

In order to estimate the influence of these sources of error, we perform a quasi-stationary simulation of a single exchange gate. We use the Hund-Mulliken approach of Reference \cite{Burk} adapted to our configuration. The quasi-stationary choice is justified by the relatively low errors expected for one single gate; however, for a larger circuit, a more accurate description of dynamical effects would be required. We consider a simplified model of a ``traveling'' quantum dot moving along the $x$ direction, e.g., a confining potential moving with constant velocity in the quantum wire and trapping an electron.

In the absence of external fields, the interaction between two electrons is described by the sum of free-particle and Coulomb Hamiltonians:
\begin{equation}
H_{orb}= \frac{1}{2m^*}( {\bf p}^2_1+ {\bf p}^2_2)+ V_c({\bf r}_1)+ V_c({\bf r}_2)+ \frac{e^2}{\kappa|{\bf r}_1-{\bf r}_2|}
\end{equation}
where $e$ is the electron charge and $\kappa$ the dielectric constant. We assume a quartic potential of confinement in the $y$-direction \cite{Burk} of the form
\be
V_c(y)= \frac{m^*\omega^2}{8} (y^2- a^2)^2
\ee
which approximates adequately the merging of two harmonic potential wells describing the free wires. Outside the gate the electrons move in a 1D quantum wire described by the Hamiltonian (\ref{H0}) with a parabolic $V_0$. 

In the $x$ direction we take the wavefunction to be a Gaussian of width $\sigma$ given by the traveling potential, which we assume for simplicity to be parabolic. The choice of $\sigma$ (instead of the usual harmonic frequency) is motivated by the fact that we are interested in the influence of the ``spreading'' $\sigma$ on the gate accuracy, having in mind that the ``traveling'' potential could be controlled experimentally (e.g., a wave propagating along the wire with controllable shape).

In the $y$ direction the electron is confined by an (experimentally fixed) parabolic potential and we assume that the corresponding wavefunction is described by the usual harmonic oscillator ground state (there is no interband coupling). Thus, the single particle state for a single quantum wire is:
\begin{equation}
\phi(x,y)= \left( \frac{m^* \omega}{\hbar\pi^2 \sigma^2}\right)^{1/4} \exp\left( -\frac{x^2}{2 \sigma^2}-\frac{y^2}{2}\frac{m^* \omega}{\hbar}\right)
\end{equation}
The geometry of the two-qubit gate is described by two parallel quantum wires situated at $y=\pm a$. Let $\phi_{-a}({\bf r})= \braket {{\bf r}} {1}$ and $\phi_{+a}({\bf r})= \braket{{\bf r}}{2}$ be the one-particle orbitals centered at $y= \mp a$, respectively. We also denote by $S= \braket{2}{1}= \int \phi^*_{+a}(x,y) \phi_{-a}(x,y) dx dy$ the overlap of the orbital wavefunctions of the two electrons.

Since we want to study the effect of non-synchronization, we assume the that each qubit propagates with a different velocity $v_i$ and can enter the gate at different times (thus there is a time-lag between the two). Hence our analysis extends the model presented in Ref.~\cite{Barnes} of a surface acoustic wave in which all qubits are perfectly synchronized.

In the Hund-Mulliken approach the {\em orbital} part of the two-particle Hilbert space has four dimensions and includes the states of double occupancy. The basis can be written as:
\begin{equation}
\Psi_{\pm a}^d({\bf r}_1,{\bf r}_2)= \Phi_{\pm a}({\bf r}_1) \Phi_{\pm a}({\bf r}_2)
\end{equation}
\begin{equation}
\Psi_{\pm }^s({\bf r}_1,{\bf r}_2)= [\Phi_{+a}({\bf r}_1)\Phi_{-a}({\bf r}_2)\pm\Phi_{-a}({\bf r}_1)\Phi_{+a}({\bf r}_2)]/\sqrt{2}
\end{equation}
where the orthonormal single-particle states are:
\begin{equation}
\Phi_{\pm a}=\frac{\phi_{\pm a}-g\phi_{\mp a}}{\sqrt{1-2Sg+g^2}}
\end{equation}
and 
\begin{equation}
g=\frac{1-\sqrt{1-S^2}}{S}
\end{equation}

The exchange coupling $J$ is calculated as the energy difference between the lowest singlet and the triplet state, obtained by diagonalizing the Hamiltonian matrix (written here in the basis described above): 
\begin{equation}
H_{orb}= \pmatrix{U_1 & X & -t_{H_1} & 0 \cr
X & U_2 & -t_{H_2} & 0 \cr
-t_{H_1} &-t_{H_2} &V_+ &0 \cr
0 &0 &0 &V_-}
\end{equation}
In is straightforward to see that $H_{i4}=0,\ i=1\ldots 3$, due to the fact that $\Psi^d_{-a}$ is antisymmetric under the exchange $1\leftrightarrow 2$, whereas the Hamiltonian and the other three basis states are symmetric. Note that compared to Reference \cite{Burk}, in our case we have $U_1\ne U_2$ and $t_{H1}\ne t_{H2}$, due to the asymmetry of the two electrons (we allow one of them to arrive before the other at the gate, if they are not synchronized).

\begin{figure}
\putfig{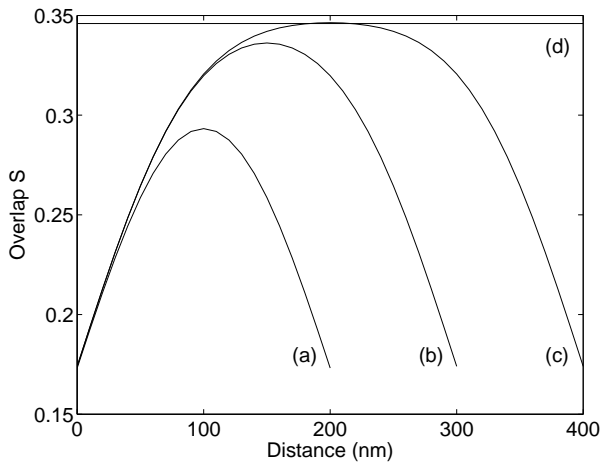}{8}
\caption{The overlap $S$ of two interacting orbitals as a function of the distance traveled along the gate. The qubits are allowed to enter the gate gradually and the gate lengths are (a) 200 nm, (b) 300 nm, (c) 400 nm. In (d) both qubits are assumed to enter the gate completely at $t=0$. The two qubit gate was taken as perfectly symmetrical ({\em i.e.} electrons enter the gate simultaneously and with the same velocity).}
\label{J_S}
\end{figure}

\begin{figure}
\putfig{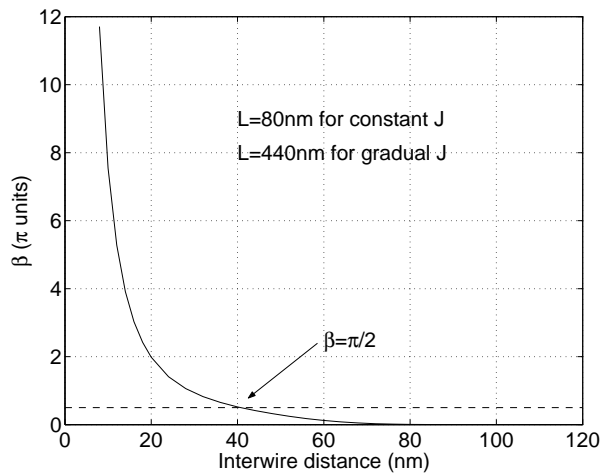}{8}
\caption{Exchange integral $\beta$ as a function of the inter-wire distance $2a$, for a symmetric two-qubit exchange gate.}
\label{J_a}
\end{figure}

\begin{figure}
\putfig{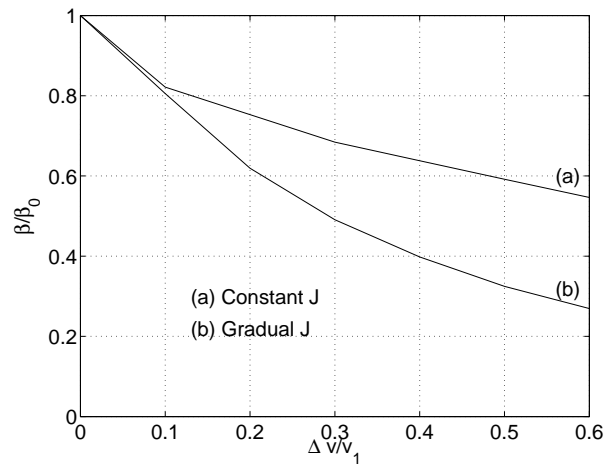}{8}
\caption{Ratio of actual and ideal values of $\beta$ as a function of the relative difference in velocities $(v_2-v_1)/v_1$ of the two electrons. The two qubits are assumed to enter the gate simultaneously and (a) completely from $t=0$ or (b) gradually.}
\label{J_v1}
\end{figure}

\begin{figure}
\putfig{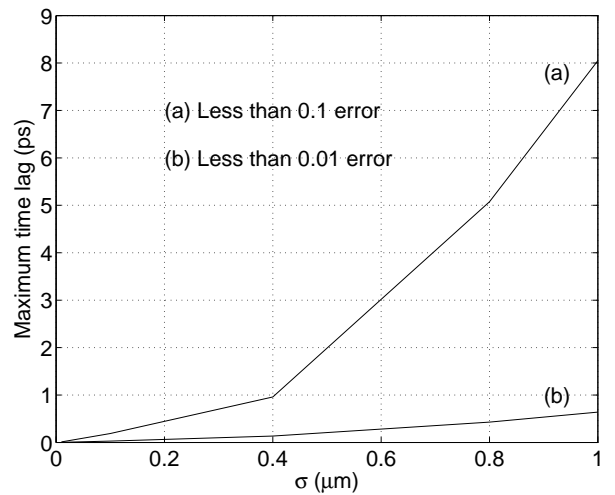}{8}
\caption{Maximum acceptable time lag between the interacting electrons for (a) less than 10\% and (b) less than 1\% error per gate. Both qubits are allowed to enter the gate gradually.}
\label{J_t0}
\end{figure}

Our results are summarized in Figures \ref{J_S}-\ref{J_t0}, for GaAs ($\kappa= 13.1$) taking $\hbar \omega= 3$\,meV, $v=10^5$\,m/s, an inter-wire distance $2a=40$\,nm and $\sigma=10^{-7}$m (unless specified otherwise). The goal is to find the gate length for a fixed value of the exchange integral, namely $\beta= \pi/2$ corresponding to the $\sf \sqrt{SWAP}$ gate. We allow the electrons to enter the gate gradually; this effect is specific to mobile qubits and, to the best of our knowledge, has not been accounted for so far. This phenomenon induces a variation in the overlap $S$ over the length of one gate (see Figure \ref{J_S} (a)-(c), compared to (d)). Its inclusion gives a more accurate description of the gate operation, especially for weakly localized electrons (i.e., for relatively large values of $\sigma$). By assuming that the exchange coupling $J$ is constant, the gate length would be drastically overestimated, and errors due to lack of symmetry/synchronization would be underestimated (see Figures \ref{J_a} and \ref{J_v1}).

As already reported in the literature, $\beta$ has an approximately exponential decay with the inter-wire distance $2a$ (see Figure \ref{J_a}). This result imposes the constraint of a tight control on the parameter $a$. Individual gate calibration is in principle possible if the wires are build through depletion of a 2DEG (see Section V for more details).

Unlike proposals using electrons trapped inside quantum dots, gates acting on mobile spins are also affected by the lack of synchronization between the interacting qubits, i.e., the two electrons can have different velocities and can enter the gate at different times. Deviations from the nominal velocity $v$ are not especially critical, as long as symmetry is preserved (in this case $\beta$ is roughly $\sim 1/v$). However, the gate accuracy is more sensitive to a difference in velocity between the two qubits. This is illustrated in Figure \ref{J_v1}, where it was assumed that the interacting electrons reach the gate simultaneously, but have different velocities. In consequence, precautions should be taken in order to prepare the qubits as highly monoenergetic, tightly synchronized electrons. 

Gate accuracy decreases even further when there is a time lag between the interacting electrons. The acceptable time lag, for a gate error less than 10\% and 1\%, is plotted in Figure \ref{J_t0}, as a function of $\sigma$ (for a gradual $J$). It is apparent from this plot that Gaussians with a large spreading $\sigma$ are preferred to more localized orbitals from the point of view of asymmetry-induced decoherence. Weakly localized orbitals have the drawback of a larger computation time, since successive electrons should be injected at longer time intervals. 

\section{Discussion}

\subsection{Programming}

The main advantage of having a mobile qubit is that one can use {\em cold programming}, i.e., all gates are set before ``launching'' the electrons. This avoids the use of ultrafast (i.e., subdecoherent) electronics for gate operations. This property is essential and represents a distinct advantage of the proposed quantum architecture over other solid-state proposals. The gating sequence needed for the proposed experiment can be pre-programmed using {\it static electric fields only}. Programming is done by switching on/off the gates and in this way any quantum algorithm can be implemented. Proposals that do not benefit from this feature can still be used for the implementation of quantum specialized circuits, which are hard-wired for only one algorithm; in that case, the loss in flexibility is compensated by improved robustness.

\begin{figure}
\putfig{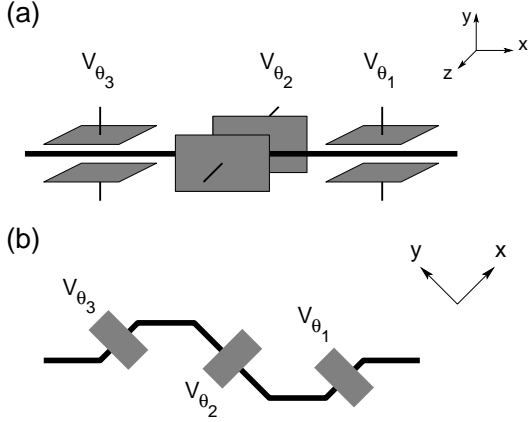}{7}
\caption{Two architectures for an arbitrary single-qubit gate $U(\theta_1,\theta_2,\theta_3)= e^{i\theta_1\sigma_z}e^{i\theta_2\sigma_y}e^{i\theta_3\sigma_z}$ using the spin-orbit interaction. $V_{\theta_i}$ are three gate potentials generating the Rashba phases $\theta_i$. (a) The electric field in the three Rashba regions is $E_y$, $E_z$ and $E_y$, respectively. This setup requires both top/bottom and lateral gates; (b) If only top/bottom gates are available (all three Rashba regions have now an $E_z$ field), the gate has a different setup: the electron is moving along three different directions $Ox$, $Oy$ and $Ox$, respectively. The gate is now $U'(\theta_1,\theta_2,\theta_3)= e^{i\theta_1\sigma_y}e^{i\theta_2\sigma_x}e^{i\theta_3\sigma_y}$; spin rotations around two perpendicular axes (in this case $Ox$ and $Oy$) are sufficient for performing any single-qubit gate.}
\label{su2}
\end{figure}

In our implementation the control of single-qubit gates amounts to switching on and off the voltage applied on a local capacitor (see Figure \ref{su2}). Since any $SU(2)$ matrix can be parametrized as $U(\theta_1,\theta_2,\theta_3)= e^{i\theta_1 \sigma_z} e^{i\theta_2 \sigma_y} e^{i\theta_3 \sigma_z}$ \cite{univ1}, an arbitrary single qubit gate can be enacted with only three Rashba active regions (see Fig.~\ref{su2}). The local control of the electric field between a maximum value (when the gate is active) and zero (gate inactive) can be achieved with the present technology \cite{Rashba1, Rashba2}.

\begin{figure}
\putfig{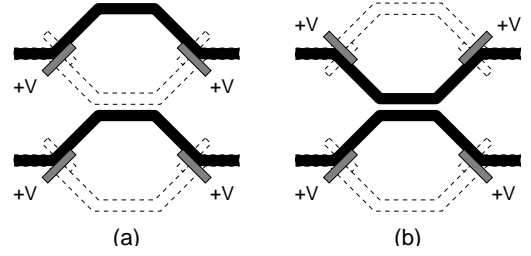}{7}
\caption{On/off switching mechanism for the two qubit gate. Each qubit has two pairs of depleting gates (grey rectangles). The electron can be directed along one of the two possible trajectories (top or bottom, marked in black) by applying a depletion voltage $+V$ on one of the two pairs of gates; no voltage is applied on the other pair shown as a dotted rectangle. When interaction is required, the top qubit is flipped to the lower trajectory state as in (b); in (a) there is no interaction between qubits. Note that the trajectory length is the same in both cases.}
\label{switch}
\end{figure}

There are a number of methods for controlling a two-qubit exchange gate. While reducing the distance between the wires in the interaction area does not offer the possibility of switching the gate off completely (and also affects synchronization), lowering the inter-wire potential when the gate is active (again by the control of a local electric field) is a more attractive option. As mentioned previously, for universality it is sufficient to have a single two-qubit gate like $\sqrt{\sf SWAP}$. In this case programming reduces to switching on and off the gate placed along the qubit path. Also, since the two qubit gate is constant, its large scale integration and the optimization of its geometry (gate length, inter-wire distance etc) are easier than for a variable gate. A switching mechanism for the two-qubit gate is presented schematically in Figure \ref{switch}. Using two pairs of depleting gates, the electron can be directed along two alternative paths, corresponding to to the active and inactive gate, respectively.

\subsection{Decoherence and scalability}

Our estimate of less than $10^3$ gates (corresponding to a spin coherence length of 100$\mu$m) falls short of the threshold required for fault tolerant quantum computation ($10^4$, according to \cite{fault_tolerant}). Progress in this direction can came from materials with longer spin coherent length or higher $\alpha$ (implying shorter gates).

Our model can also rely on several results from encoded universality using the exchange interaction (see, for instance, \cite{ex_DFS}) to fight particular types of errors, albeit at the expense of an increased gate overhead.

It should be mentioned that in the simulations presented above we did not consider the time spreading $\sigma(t)$ of the Gaussian, assuming tacitly that the ``traveling'' potential remains constant during propagation. If this effect is also taken into account, then it could limit the maximum number of two-qubit gates to less than 100. If a larger number of gates become technologically possible and required for the particular application considered, extra ``reshaping'' steps should be added. Considerable improvement in the maximum number of gates can come from putting the two wires into closer contact, i.e., by reducing the inter-wire distance $2a$, since this implies shorter gate lengths.

\section{Implementations}

The proposal described here can benefit from recent advances in nanofabrication. High density arrays of parallel nanowires with diameters as small as 8 nm and center-to-center distances of 16 nm were fabricated \cite{melosh}. This techniques is applicable to both metallic and semiconductor nanowires and moreover, simple circuits of crossed nanowires were produced (with junction densities larger that $10^{11}\,{\rm cm}^{-2}$).

\subsection{Semiconductor heterostructures}

A large number of experimental results in spintronics are obtained for GaAs, InAs and related heterostructures. Spin injection and spin transport have been demonstrated primarily for these materials, therefore it is natural to consider them as first candidates for the implementation of our proposal.

One of the technological alternatives that can prove suitable for implementing our proposal is based on the depletion of a 2DEG using surface gates (see for instance \cite{LossDiVi}). Gate regions can be patterned by extending the depletion region created by a split gate so that a semiconductor island is defined only in the active area. This approach has the very important advantage of allowing individual calibration of each gate by varying the length and width so that the desired rotation angle is obtained. 

\subsection{Silicon}

One of the reasons why Si is a good candidate for implementing the present model is its experimentally demonstrated suitability for building low-dimensional systems. In particular, purely ``zero-dimensional'' structures have been produced in Silicon-on-Insulator (SOI) substrates \cite{SOI2}. Since the active area of these substrates consists in a very thin Si layer sandwiched between two oxide layers, the structures fabricated in this technology are naturally confined in two dimensions. Thus, our proposal can be implemented by patterning quantum channels into the thin Si layer that acts as a 2DEG, either by confining potentials (this solution has the advantage of a conceptually straightforward manipulation of the exchange interaction, but introduces an extra term in the Rashba Hamiltonian), or by shallow oxide trenches. Moreover, a multiple sandwiched structure of Si and SiO$_{2}$ fabricated using the bonding technology \cite{SOI2} and suitable for a three-dimensional arrangement of the quantum wires can be envisaged. This option will relax one of the constraints imposed by our model, where only two-qubit gates between neighboring qubits can be directly implemented. If two-qubit interactions are possible between any two qubits of the circuit, the number of total operation required for one computation can be considerably reduced.

However, the issues of state preparation and measurement, the coherence length of mobile spins and the spin-orbit interaction in Si have to be addressed separately, in order to achieve the same level of accuracy in description as for compound semiconductors. The authors have no knowledge of a definite answer to the question of spin-coherent transport in Si; nevertheless, the transverse relaxation time of electron spin in Si samples at $1.4$\,K was found to be of the order of hundreds of \,$\mu$s \cite{Si_decoherence_1}, while the phonon-induced decoherence for localised electrons in Si was predicted to be negligible \cite{Si_decoherence_2}. Those are promising results for future investigation of the coherence length of electron spin in Si one-dimensional systems, and if confirmed by further investigations might recommend this material, particularly in the SOI technology, as a good option for the fabrication of quantum circuits. 

A special attention has to be paid to the exchange coupling in Si, since anisotropy and an oscillatory behaviour can be observed, due to intervalley interactions in the six minima of the conduction band \cite{Vrij, Si_exchange_1}. However, it was shown in \cite{Si_exchange_2} that these perturbations are reduced for symmetric quantum wells and for strained structures, where one conduction band valley becomes dominant. Therefore, control of the anisotropy in exchange can be expected to be technologically viable in the confined 1D systems we employ here. 

\subsection{Carbon Nanotubes (CNTs)}

The main attractiveness of CNTs consists in their potential of behaving as ballistic conductors at room temperature. Tentative studies of spin transport in CNTs have been reported in \cite{CNT_transport_1, CNT_transport_2, CNT_transport_3}; if the results obtained for charge transport (typical coherence lengths of 1\,$\mu$m at room temperature) are reproduced for spin transport, CNTs could emerge as one of the most promising supports for solid-state quantum computing with mobile spin qubits. Particular attention should be paid not only to the study of single spin transport, but also to the suitability of CNTs for entanglement transport. In general, the decoherence time of a spin singlet (each of the electrons in a separate quantum dot or channel) can be assimilated with the decoherence time of a single spin \cite{Loss_pc}. However, it was demonstrated in \cite{LL_singlet} that if Luttinger liquid (LL) interactions are taken into consideration, the decoherence time of such a singlet becomes almost zero. Since CNTs are better described by the LL model (see for instance \cite{CNT_LL}), this effect is a potential limit for the performances of quantum computing with mobile spin qubits in CNTs, and experimental results to confirm or infirm these considerations would be of great interest. 

Regarding the interactions required for an universal set of qubit transformations, it was predicted that spin-orbit coupling exists in CNTs, and moreover, it does not interfere with the spin-charge separation \cite{CNT_Rashba}; the exchange coupling was discussed in \cite{CNT_transport_2}. For the injection step, a specific spin pump for Luttinger liquids was proposed in \cite{LL_pump}. More theoretical and experimental results are expected, in order to fully assess the suitability of CNTs for the implementation of quantum computing with mobile spin qubits. 

\section{Conclusions}

In this article we have described and assessed a novel, all-electrical model for quantum computing using mobile spin qubits in quantum one-dimensional systems. Besides benefiting from all advantages characteristic to a solid-state environment, our proposal combines the processing and transport of quantum information in a more natural way. Therefore, it eliminates the requirement of ultrafast switching of local electric and magnetic fields common to static qubit proposals. Solely static electric fields are used for the manipulation of quantum information, and the fidelity of individual transformations in a specific material depends only on an accurate control of the gate length (for one-qubit gates), respectively of the gate length and qubit energy and synchronisation for two-qubit gates. Moreover, the gates are programmable (they can be turned on and off by external potentials), therefore universal quantum computing, as opposed to dedicated circuits, is in principle possible.

We have estimated the influence of various sources of error on individual two-qubit gates, via a quasi-stationary Hund-Mulliken approach, in a realistic setting allowing for time variations of the exchange coupling $J$. It was found that the main trade-off in the flying qubit approach is the requirement of synchronizing the qubits. It is thus essential to have highly monoenergetic electrons launched simultaneously. This can be accomplished by properly tailored energy filters and synchronized single-electron injectors at the preparation stage.

However, the drawback of dealing with extra sources of decoherence is compensated by an improved flexibility in the implementation of quantum gates, where dynamic effects such as the spin-orbit interaction can be used. The experimentally proved, non-dispersive Rashba effect offers a very promising alternative for the implementation of one-qubit gates. The steps of qubit preparation and measurement also benefit from increased flexibility, with a variety of available methods. It is particularly encouraging to note that the qubit measurement is much simplified compared to the case of static qubits. Any further advances in spin transport, spin injection and the design of spin filters can be naturally incorporated in our scheme, with the effect of improving the overall accuracy of the computation. 

As with all the results in spintronics and quantum information processing, further increase in the control of errors and correspondingly in the maximum number of operation that can be performed during one computation depends on future technological advances. Our proposal provides a conceptually simple and flexible framework for the demonstration of quantum computing in several materials with large spin coherence length that can be processed into parallel one-dimensional wires, such as compound semiconductors, heterostructures, silicon-on-insulator (SOI) or CNTs.

\begin{acknowledgments}
The authors are grateful to I.~D'Amico, P.~Zanardi, E.~Pazy and F.~Udrea for useful discussions. A.E.P. wishes to thank Gonville and Caius College, Cambridge for awarding her a Research Fellowship during which this study was performed. 
\end{acknowledgments}

\end{document}